\documentclass[aip,apl,amsmath,amssymb,reprint,numerical,floatfix,superscriptaddress]{revtex4-2}
\usepackage{graphicx}
\graphicspath{{./images/}}
\usepackage{dcolumn}
\usepackage{bm}% bold math
\usepackage{mathtools}
\usepackage[caption=false]{subfig}
\usepackage{amssymb}
\expandafter\let\csname equation*\endcsname\relax
\expandafter\let\csname endequation*\endcsname\relax
\usepackage{amsmath}
\usepackage{amsbsy}
\usepackage{booktabs}
\usepackage{verbatim}
\usepackage{array}
\newcolumntype{P}[1]{>{\centering\arraybackslash}p{#1}}
\usepackage[colorlinks=true,citecolor=blue]{hyperref}

\AtBeginDocument{}%

\begin{document}

% Use the \preprint command to place your local institutional report
% number in the upper right-hand corner of the title page in preprint mode.
% Multiple \preprint commands are allowed.
% Use the 'preprintnumbers' class option to override journal defaults
% to display numbers if necessary
%\preprint{}

%Title of paper
\title{Thermalization of radiation-induced electrons in wide-bandgap materials: A first-principles approach}

% repeat the \author .. \affiliation  etc. as needed
% \email, \thanks, \homepage, \altaffiliation all apply to the current
% author. Explanatory text should go in the []'s, actual e-mail
% address or url should go in the {}'s for \email and \homepage.
% Please use the appropriate macro for each type of information

%-------------------------------------------------------------------------------------------------------------%
%  Authors
%-------------------------------------------------------------------------------------------------------------%
% \affiliation command applies to all authors since the last
% \affiliation command. The \affiliation command should follow the
% other information
% \affiliation can be followed by \email, \homepage, \thanks as well.
\author{Dallin O. Nielsen}
%\email{dallin.nielsen@utdallas.edu}
%\homepage[]{Your web page}
%\thanks{}
%\altaffiliation{}
\address{The University of Texas at Dallas, Department of Materials Science and Engineering, 800 W. Campbell Rd., Richardson, Texas 75080, USA}

%\author{Chris G. Van de Walle}
%\email{vandewal@ucsb.edu}
%\address{University of California, Santa Barbara, Materials Department, 2510 Engineering II, Santa Barbara, CA 93106-5050, USA}

%\author{Sokrates T. Pantelides}
%\email{pantelides@vanderbilt.edu}
%\address{Vanderbilt University, Department of Physics and Astronomy, Vanderbilt University, Nashville, TN 37235-1824, USA}

%\author{Ronald D. Schrimpf}
%\email{ron.schrimpf@vanderbilt.edu}
%\address{Vanderbilt University, Department of Electrical and Computer Engineering, VU Station B 351824, 2301 Vanderbilt Place, Nashville, TN 37235-1824, USA}

%\author{Daniel M. Fleetwood}
%\email{dan.fleetwood@vanderbilt.edu}
%\address{Vanderbilt University, Department of Electrical and Computer Engineering, VU Station B 351824, 2301 Vanderbilt Place, Nashville, TN 37235-1824, USA}

\author{Massimo V. Fischetti}
%\email{max.fischetti@utdallas.edu}
\address{The University of Texas at Dallas, Department of Materials Science and Engineering, 800 W. Campbell Rd., Richardson, Texas 75080, USA}

%Collaboration name if desired (requires use of superscriptaddress
%option in \documentclass). \noaffiliation is required (may also be
%used with the \author command).
%\collaboration can be followed by \email, \homepage, \thanks as well.
%\collaboration{}
%\noaffiliation

\date{\today}

\begin{abstract}
The present study is concerned with simulating the thermalization of high-energy charge carriers (electrons and/or electron-hole pairs), generated by ionizing radiation, in diamond and $\beta$-Ga$_2$O$_3$. Computational tools developed by the nuclear/particle physics and electronic device communities allow for accurate simulation of charge-carrier transport and thermalization in the high-energy (exceeding $\sim$~100~eV) and low-energy (below $\sim$~10~eV) regimes, respectively. Between these energy regimes, there is an intermediate energy range of about 10-100 eV, which we call the ``10-100 eV gap'', in which the energy-loss processes are historically not well-studied or understood. To close this ``gap'', we use a first-principles approach (density functional theory) to calculate the band structure of diamond and $\beta$-Ga$_2$O$_3$ up to $\sim$~100~eV along with the phonon dispersion, carrier-phonon matrix elements, and dynamic dielectric function. Additionally, using first-order perturbation theory (Fermi's Golden Rule/first Born approximation), we calculate the carrier-phonon scattering rates and the carrier energy-loss rates (impact ionization and plasmon scattering). With these data, we simulate the thermalization of 100-eV electrons and the generated electron-hole pairs by solving the semiclassical Boltzmann transport equation using Monte Carlo techniques. We find that electron thermalization is complete within $\sim0.4$ and $\sim1.0$~ps for diamond and $\beta$-Ga$_2$O$_3$, respectively, while holes thermalize within $\sim0.5$~ps for both. We also calculate electron-hole pair creation energies of 12.87 and 11.24~eV, respectively.
\end{abstract}

% insert suggested keywords - APS authors don't need to do this
%\keywords{}

%\maketitle must follow title, authors, abstract, and keywords
\maketitle

In electronic devices, ionizing radiation may excite charge carriers (electrons and/or electron-hole pairs) to high energies, which may cause effects ranging from transient upset to catastrophic device failure. The study of this issue is vital to space exploration and other applications where radiation interacts with electronics. To protect these devices, the nuclear/particle physics community has developed computational tools~\cite{Reed15,Agostinelli03,Allison06,Biersack80} to simulate the thermalization/transport of these excited carriers, allowing one to adapt a given device to mitigate the potentially destructive effects. These codes accurately simulate the thermalization process down to $\sim$~100~eV. The electronic device community has studied extensively charge-carrier transport for lower energies, but only up to $\sim$~10~eV~\cite{MVF88,Fang19,Ghosh17,Bertazzi09,reaz_2021}. In the intermediate energy range of about 10-100~eV, which we call the ``10-100 eV gap'', the energy-loss processes are historically not well-studied. In another paper~\cite{nielsen_2023}, we have reported on our work to close this ``gap'', using a first-principles approach. While this previous paper focuses on charge-carrier thermalization in GaN, here, we employ the same techniques for diamond (C) and $\beta$-Ga$_2$O$_3$.

Theoretical work on electronic transport in this intermediate energy range has been published in the past. Notably, in 1956, Pines~\cite{Pines_1956} showed that energy losses in this regime are primarily due to valence-electron plasmon emission. Following this assumption, others have studied the energy-loss processes in the ``10-100 eV gap'' for various materials, including phosphors~\cite{rothwarf_1973,kingsley_1970}, semiconductors~\cite{alig_1980}, SiO$_2$~\cite{ausman_1975}, and scintillators~\cite{prange_2015,prange_2017}. In most of these works, however, the authors utilized the free-electron model to calculate the electronic band structure along with semi-empirical matrix elements and/or scattering rates or a quasi-free gapped band structure and the Callaway-Tosatti model energy loss function~\cite{prange_2015,prange_2017,callaway_1959,tosatti_1971}. Only the electron-phonon scattering rates were computed using {\it ab initio} methods up to 100 eV by Prange {\it et al.}~\cite{prange_2015}. 

In this and our previous paper, we employ an {\it ab initio} approach to calculate material properties (electronic band structure, phonon dispersion, dynamic dielectric function, and carrier-phonon matrix elements) using the density functional theory (DFT) package Quantum ESPRESSO~\cite{giannozzi_2009} (QE). Additionally, using first-order perturbation theory (Fermi's Golden Rule/first Born approximation), we calculate the relevant charge-carrier scattering rates, including plasmon emission, impact ionization, and phonon scattering. Finally, we employ full-band Monte Carlo (MC) techniques~\cite{MVF88,nielsen_2023} to solve the semi-empirical Boltzmann transport equation to simulate the full thermalization of 100-eV electrons and the generated electron-hole pairs.

Wide- and ultrawide-bandgap materials [including C (E$_g=5.47$~eV) and  $\beta$-Ga$_2$O$_3$ (E$_g=4.8$~eV)] have received much attention in the study of ionizing radiation effects in electronics~\cite{Akturk17,Witulski18,landstrass_1990,campbell_2000,ma_2023,Fleetwood22,ives_2015,jiang_2017} due to their potentially radiation-resistant properties. In particular, a wide band gap entails a larger electron-hole pair creation energy, which leads to fewer excited pairs and potentially less damage. In addition to their importance in the study of radiation effects, such materials are especially well-suited for high-power electronics applications due to their relatively high breakdown fields ($\beta$-Ga$_2$O$_3$: 6-8~MV/cm, C: 7.7-10~MV/cm). In such applications, the study of charge-carrier transport in the ``10-100 eV gap'' is important as fields are often high enough that carriers find themselves in this regime.

We note that in this letter we restrict the scope of our work to a low-dose rate of irradiation. This restriction entails a low density of carriers, permitting us to ignore both the radiation-induced heating of the crystal as well as short-range carrier-carrier scattering and plasmon absorption, as the number of generated plasmons is assumed to be low enough for the distribution to remain at thermal equilibrium.

We start by performing a relaxation of the crystal structures of C and $\beta$-Ga$_2$O$_3$, using the {\it vc-relax} function in QE. For the band structure of C, we employ a norm-conserving pseudopotential~\cite{Schlipf15} with PBE~\cite{Perdew96} exchange-correlation (XC) functionals and Heyd, Scuseria, and Ernzerhoff (HSE06)~\cite{heyd_2003,heyd_2006} hybrid functionals. These hybrid functionals mix the PBE XC potential with a fraction of exact exchange from Hartree-Fock theory. We use also a norm-conserving pseudopotential with PBE XC functionals for $\beta$-Ga$_2$O$_3$. In the case of $\beta$-Ga$_2$O$_3$, however, we do not use a hybrid functional approach due to the computational cost in calculating electronic bands up to $\sim100$~eV above the conduction band edge. In the self-consistent calculations for both, we use a uniform $8\times8\times8$ {\bf k}-point grid. We employ a fraction of exact exchange of 0.11, for C, and plane-wave cutoff energies of 120 and 100~Ry, for C and $\beta$-Ga$_2$O$_3$, respectively.

The unit cells of C and $\beta$-Ga$_2$O$_3$ are face-centered cubic with a space group of Fd3m and base-centered monoclinic with space group C2/m, respectively. Following the relaxations, we obtain lattice constants of $a_0=3.567~{\rm \AA}$ for C, and $a_0=12.04~{\rm \AA}$, $b_0=3.01~{\rm \AA}$, $c_0=5.75~{\rm \AA}$, and $\beta=103.73$\textdegree for $\beta$-Ga$_2$O$_3$. 

To reach energies of $\sim100$~eV above the conduction band edge, we calculate the band structure for a total of 45 bands (4 valence bands and 41 conduction bands) in C, and 350 bands (44 valence bands, with the {\it d}-electron bands included, and 306 conduction bands) in $\beta$-Ga$_2$O$_3$. The resulting primary energy gaps are 5.45~eV (C), which nearly matches experimental results exactly, and 2.63~eV ($\beta$-Ga$_2$O$_3$), which underestimates the gap by nearly one-half. This underestimation of the gap for $\beta$-Ga$_2$O$_3$ is a well-known problem of DFT, which can be corrected with the use of HSE06 hybrid functionals~\cite{peelaers_2015}. In this work, however, we simply employ a scissors operation, shifting the conduction bands up by 2.17~eV. The full-band MC technique requires knowledge of the band structure at any given point in the Brillouin zone (BZ). We, therefore, calculate the band structure for a set of points (C: 1661, $\beta$-Ga$_2$O$_3$: 3614) spanning the irreducible wedge of each material, which can be used to interpolate $E_n({\bf k})$ for any point {\bf k} in band $n$.

To account for the carrier-phonon interaction, we evaluate the phonon dispersion using density functional perturbation theory (DFPT) in QE. As QE does not currently support hybrid functional calculations for this and the remaining first-principles calculations, from here on, we use a USPP~\cite{vanderbilt_1990} pseudopotential with PBESOL~\cite{perdew_2008,perdew_2009} XC functionals for C. We calculate the dispersion for the same set of points in the irreducible wedge for the interpolation in the MC code. Subsequently, we calculate the carrier-phonon matrix elements on uniform $8\times8\times8$ {\bf k}- and {\bf q}-point grids via the code EPW (electron-phonon coupling using Wannier functions)~\cite{Ponce16,Giustino07}. 

We now turn to the evaluation of the scattering rates. We calculate the carrier-phonon scattering rates, using Fermi's golden rule:
\begin{figure}[b]
\includegraphics[height=3.0in, width=3.375in]{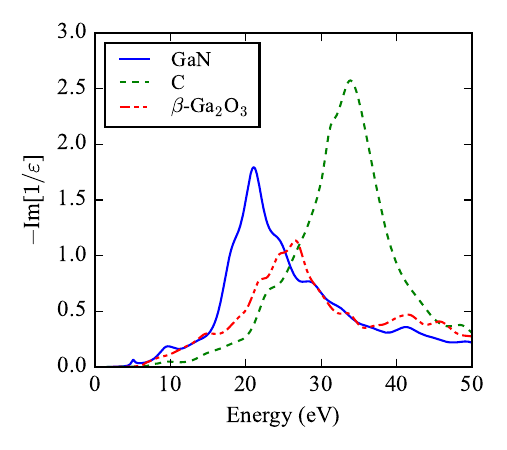}
\caption{\label{fig:loss_functions} The electron energy-loss functions of GaN, C, and $\beta$-Ga$_2$O$_3$, calculated using the turboEELS code within QE.}
\end{figure}
\begin{multline}
\frac{1}{\tau_{\eta}({\bf k},n)} \approx \frac{2\pi}{\hbar} \sum_{n',{\bf q}}\left| 
    g_{nn'}^{\eta}({\bf k},{\bf k}') \right|^{2} \left( N_{\bf q}+\frac{1}{2}\mp\frac{1}{2} \right ) \\
           \times \delta\left[E_n({\bf k})-E_{n'}({\bf k}')\pm\hbar\omega_{\bf q}^\eta\right],
\label{eq:FGR_EPW}
\end{multline}
with
\begin{equation}
N_{\bf q}=\frac{1}{e^{(\hbar\omega_{\bf q}/k_{\rm B}T)}-1}.
\label{eq:bose_einst}
\end{equation}
Here, {\bf k} represents the initial wave vector of a carrier in band $n$, scattering to a final state (${\bf k}'$,$n'$) via a phonon of wave vector {\bf q}, frequency $\omega$, and branch $\eta$, and $g_{nn'}^{\eta}({\bf k},{\bf k}')$ are the carrier-phonon matrix elements from EPW. To evaluate the summation over the delta function, we employ a technique similar to that used by Fischetti and Laux~\cite{MVF88} with the notable exception that here, we use  Bl{\"o}chl's tetrahedron method~\cite{Blochl94} to evaluate the density of states. Further details can be found in our previous paper~\cite{nielsen_2023}.
\begin{figure*}
\includegraphics[height=4.0in, width=6.75in]{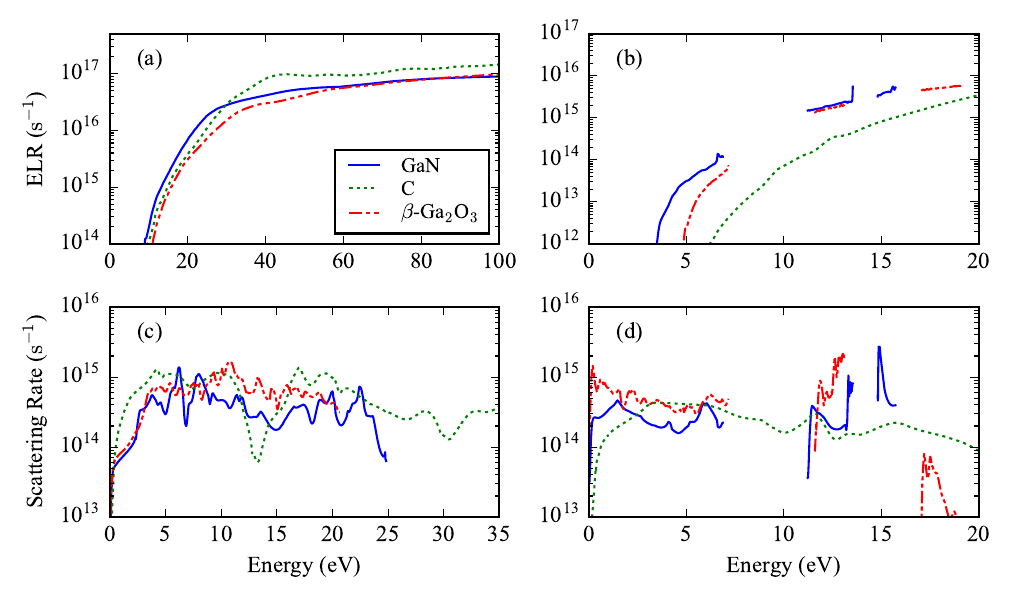}
\caption{\label{fig:scattering_rates} (a) Electron energy-loss rates, (b) hole energy-loss rates, (c) electron-phonon scattering rates, and (d) hole-phonon scattering rates for GaN, C, and $\beta$-Ga$_2$O$_3$. The legend appearing in frame (a) applies to all frames.}
\end{figure*}

Information on impact ionization and plasmon emission can be found in valence-electron energy-loss spectra and the related electron energy-loss function (${\rm Im}\left[{-1}/{\varepsilon({\bf q},\omega)}\right]$). The majority of the features (peaks and shoulders) in these spectra represent single-particle excitations (impact ionization) of the valence electrons into the conduction states, leaving holes in the valence states [electron-hole pairs (EHPs)]. The remaining peaks represent collective excitations of the valence electrons (plasmons), which decay via Landau damping~\cite{Landau46} into EHPs. Fermi's golden rule and the dissipation-fluctuation theorem can be used to obtain a scattering-rate equation for both impact ionization and plasmon emission [the energy-loss rate (ELR)]:
\begin{multline}
\frac{1}{\tau_{n}^{\rm (ELR)}({\bf k})} = 
   \frac{2\pi}{\hbar} \sum_{n'} \int\frac{{\rm d}{\bf q}}{(2\pi)^3} \frac{e^{2}\hbar}{q^2} 
       \int\frac{{\rm d}\omega}{2\pi} \ {\rm Im} \left[\frac{-1}{\varepsilon({\bf q},\omega)}\right] \\
           \times \delta \left [ E_n({\bf k})-E_{n'}({\bf k}+{\bf q}) \pm \hbar\omega \right]. 
\label{eq:ELR}
\end{multline}

We calculate the electron energy-loss function using the time-dependent DFT code turboEELS~\cite{Timrov15}. It takes, as an input, the desired momentum transfer ({\bf q}) in the excitation process as well as a range of energies ($\hbar\omega$). Ideally, one would repeat the calculation for a set of {\bf q} spanning the irreducible wedge of the BZ. Due to the computational cost, however, we calculate the loss function for several points along the first reciprocal-lattice vector of the BZ and assume isotropy. We have previously shown that for GaN, the ELRs are not significantly affected by the anisotropy of the dielectric function~\cite{nielsen_2023}. We conclude, therefore, that this simplification is reasonable.

We show the energy-loss functions for GaN, C, and $\beta$-Ga$_2$O$_3$ in Fig.~\ref{fig:loss_functions}. We note that these calculations do not account for excitonic effects (which would be impractical for such a large energy range). For GaN and $\beta$-Ga$_2$O$_3$ this is reasonable~\cite{fares_2019,sun_2018}, given their relatively large static dielectric constants, but for C, it is well documented that excitonic effects on the dielectric function are significant~\cite{gao_2015,kootstra_2000,phillip_1964}. To account for this effect, we fit the loss function peaks of C to EELS data~\cite{canas_2018}. We observe that the primary plasmon peaks occur at approximately 21.0, 26.5, and 33.8~eV for GaN, $\beta$-Ga$_2$O$_3$, and C, respectively. We also note that the magnitude of and area under the peak are significantly larger for C, suggesting greater ELRs (see Eq. \ref{eq:ELR}). Additionally, one may expect that fewer EHPs will be generated during the thermalization process in C, as the energy gap is larger and the plasmon energy is significantly higher, leading to larger pair-generation energies.

We plot in Fig.~\ref{fig:scattering_rates} the electron and hole ELRs [(a) and (b), respectively] and the electron- and hole-phonon scattering rates [(c) and (d), respectively]. In Fig.~\ref{fig:scattering_rates}(a), as expected, the C electron ELRs reach larger magnitudes for energies above $\sim30$~eV due to the loss function magnitude, as discussed above. We see that for each material the electron ELRs flatten as the energy increases and reach magnitudes of order $10^{17}$ s$^{-1}$. This result is in agreement with Quinn and Ferrel \cite{Quinn58, Quinn62}, Pines \cite{Pines_1956}, Penn \cite{Penn87}, and others who have calculated ELRs of similar magnitude. With such large ELRs, one may question the validity of the first Born approximation in this calculation. We and the authors mentioned above have concluded that where the ELRs are high, the electron energies are large enough to render the broadening of the electronic states acceptably small, justifying the use of perturbation theory.

By comparing the electron-phonon scattering rates [Fig.~\ref{fig:scattering_rates}(c)] with the electron ELRs, we see that phonon scattering dominates for low energies up to $\sim$~10-15~eV, depending on the material. Above this point, the ELRs rise to 1-2 orders of magnitude larger than the phonon scattering rates, rendering phonon scattering irrelevant at high energies (above $\sim15$-20~eV).

In the case of hole scattering, we see, again, a flattening of the ELRs as the energy increases with a maximum rate between $10^{15}$ and $10^{16}$ s$^{-1}$. As holes in the valence bands of these materials do not reach energies exceeding the plasma energy, this rate is lower than for electrons. In GaN and $\beta$-Ga$_2$O$_3$ we observe gaps in the rates as a result of energy gaps in the valence band structure. In both materials, the deeper bands primarily represent {\it d}-electron bands. We note also that hole-phonon scattering dominates for energies below $\sim10$~eV, and the ELRs dominate above this point. For energies greater than $\sim15$~eV, the hole-phonon scattering may be ignored in both C and $\beta$-Ga$_2$O$_3$. For GaN, however, hole-phonon scattering remains relevant for all hole energies.

Having determined the material properties and scattering rates, we develop a full-band MC code, following the work of Jacoboni and Reggiani~\cite{Jacoboni83} and that of Fischetti and Laux~\cite{MVF88}; details are given in our previous work on GaN~\cite{nielsen_2023}. We begin by defining a number of electrons (1000, here) with initial kinetic energies given by a Gaussian distribution centered at 100 eV. We apply no field, and the temperature is taken to be 300K. We then allow the electrons to move according to the equations of motion and scatter according to the calculated rates. No holes are included in the initial configuration, but we allow them to build up as the thermalization progresses.
\begin{figure}[!]
\includegraphics[height=6.0in, width=3.375in]{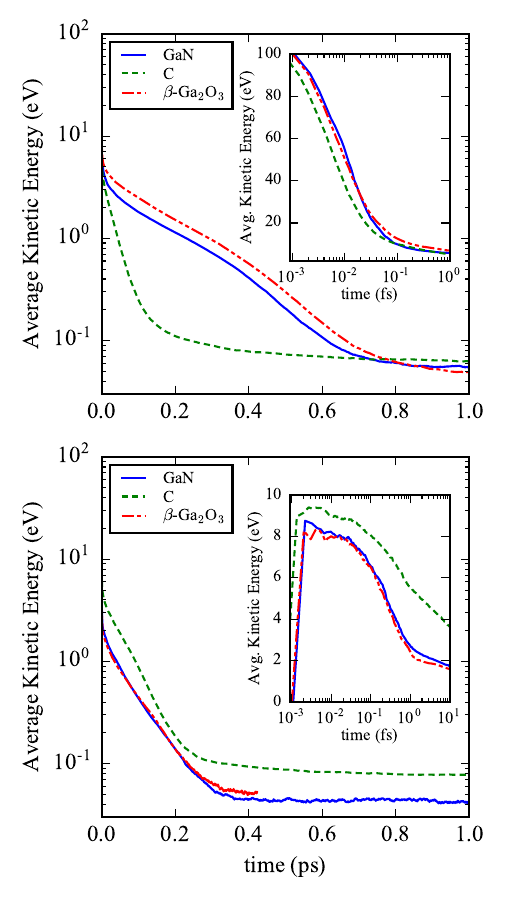}
\caption{\label{fig:thermalization} Average electron (top) and hole (bottom) kinetic energies as a function of time for GaN, C, and $\beta$-Ga$_2$O$_3$.}
\end{figure}
\begin{figure*}
\includegraphics[height=2.5in, width=6.75in]{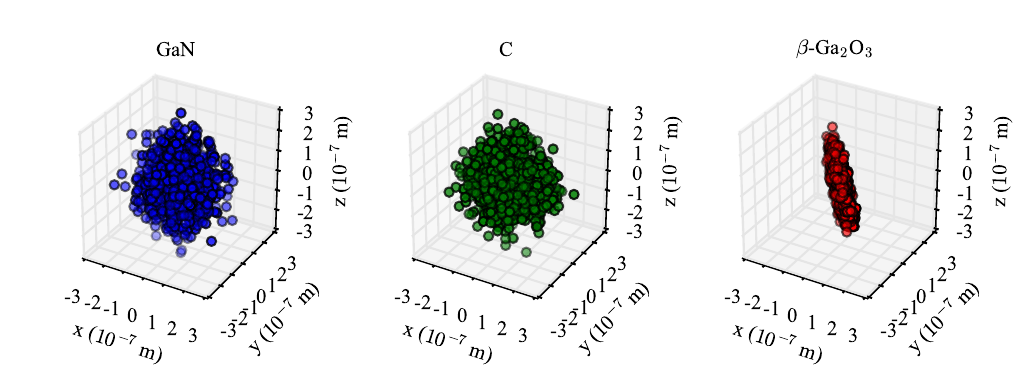}
\caption{\label{fig:real_space_position} Real-space positions of the simulated electrons after $\sim1$~ps in GaN, C, and $\beta$-Ga$_2$O$_3$ (as labeled).}
\end{figure*}

In Fig.~\ref{fig:thermalization}, we show the average electron (top) and hole (bottom) kinetic energies throughout the simulation. The larger plots of the top and bottom frames show the full thermalization process (1~ps). As the initial stages of the process are not visible in these plots, we include inset figures, showing only the first femtosecond with a logarithmic scale on the x-axis. The electron energy drops rapidly during the first femtosecond, as approximately 90\% of the electron energy is lost within 0.1~fs for all three materials. In C, due to the higher ELRs, the electrons lose energy slightly faster than those in GaN and $\beta$-Ga$_2$O$_3$ for $t<0.1$~fs. Leading up to $\sim0.1$~fs, the thermalization rate of C slows significantly, followed shortly by those of GaN and $\beta$-Ga$_2$O$_3$. The considerable decrease in the thermalization rate at this point is a result of the average electron energy falling below the plasmon peak energy. The relatively high peak energy of C is what leads to an earlier reduction in the thermalization rate. Beyond this point impact ionization and phonon emission dominate, and eventually impact ionization dissipates (when the tail of the energy distribution falls below $\sim10$-15~eV), leaving phonon emission as the primary energy-loss mechanism for the remainder of the simulation.

The larger plot of the top frame of Fig.~\ref{fig:thermalization} shows that full electron thermalization is complete in $\sim0.4$~ps for C and $\sim1.0$~ps for GaN and $\beta$-Ga$_2$O$_3$. The electrons in C thermalize more quickly due to their higher phonon scattering rates for energies below $\sim6$~eV.

For holes (Fig.~\ref{fig:thermalization}, bottom frame), we see an increase in energy at the beginning (inset figure) as a result of hole buildup, due to EHP generation. As the maximum hole energy falls below the plasmon peak energy in all three materials, the energy losses are primarily due to impact ionization and phonon emission throughout the hole thermalization. Holes for all three materials thermalize by $\sim0.5$~ps. However, the average hole energy of C is not completely thermal by this time, and it continues to drop at a slow rate beyond 1~ps. This incomplete thermalization is due to the sharp decrease in the hole-phonon scattering rate below $\sim1$~eV [see Fig.~\ref{fig:scattering_rates}(d)]. For the primary purpose of this work (thermalization in the ``10-100 eV gap''), we can ignore this issue, as it occurs for energies well below the ``10-100 eV gap''.

In addition to the thermalization rate, we also collect the real-space positions of the electrons as a function of time. For this simulation, we are interested primarily in the average displacement, so we begin with all electrons at the origin to observe the spread over time.  In Fig.~\ref{fig:real_space_position}, we plot these positions for each material after $\sim1$~ps. By this time, the electrons have traveled an average distance on the order of $\sim100$~nm in all directions for C and GaN, and primarily along the z-axis for $\beta$-Ga$_2$O$_3$. As noted in our previous paper~\cite{nielsen_2023}, this finding suggests that the charge carriers may travel through several devices, which is in agreement with Weller {\it et al.}~\cite{weller_2004}, who found that this must be true for electronic equilibrium as a condition for proper device simulation and testing.

We note also that while C and GaN appear to yield mostly isotropic spreads, the electron spread for $\beta$-Ga$_2$O$_3$ is markedly anisotropic. The ellipsoidal shape shown in Fig.~\ref{fig:real_space_position} for $\beta$-Ga$_2$O$_3$ suggests that electrons move more easily along a direction that is nearly aligned with the z-axis. Looking at the unit cell of $\beta$-Ga$_2$O$_3$ as published by Peelaers {\it et al.}~\cite{peelaers_2015}, this direction seems to align with the long edge of the parallelepiped.

Lastly, we calculate the average EHP generation energy. For C, we obtain a value of 12.87~eV/pair, which is in excellent agreement with the available experimental data: $13.1\pm0.2$~eV, using alpha particles~\cite{kozlov_1975,canali_1979},  $13.25\pm0.50$~eV~\cite{keister_2009} and  $13.05\pm0.20$~eV~\cite{morse_2007}, using continuous X-ray beams, and  $12.82\pm0.13$~eV, using a transmission X-ray detector~\cite{keister_2018}. For $\beta$-Ga$_2$O$_3$, we calculate a value of 11.24~eV/pair. We found only one published ``experimental estimation'' of the EHP creation energy in $\beta$-Ga$_2$O$_3$ by Yakimov {\it et al.}, in which a value of 15.6~eV is reported~\cite{Yakimov21}. Their method includes electron-beam-induced current measurements on a Schottky barrier mixed with an MC simulation. It is clear from Fig. 1 of their letter that for GaN and other materials with even smaller energy gaps, this method produces results that are consistent with empirical expressions~\cite{Klein68}. As the gap energy increases, the divergence among the empirical expressions increases and their agreement with experimental evidence worsens. Indeed, while for Si, Yakimov {\it et el.} and these empirical expressions were able to produce results matching experimental data, for GaN, they were all larger than the reported value of 8.9~eV~\cite{Sellin06} (a value in agreement with our results~\cite{nielsen_2023}). In the case of C, Klein's expression~\cite{Klein68} gives a value of 15.92 eV, which is significantly larger than our result and the experimental data above. Overall, it seems that as the gap energy increases, the overestimation of the EHP creation energy by the empirical expressions becomes more severe. It is reasonable, then, to expect that this will be true also for $\beta$-Ga$_2$O$_3$, for which Klein's expression gives a value of 14.04 eV. As our results for C and $\beta$-Ga$_2$O$_3$ are both $\sim22$\% smaller than those from Klein's expression, we conclude that our EHP creation energy for $\beta$-Ga$_2$O$_3$ is reasonable.

We have presented a first-principles calculation of material properties and scattering rates in C and $\beta$-Ga$_2$O$_3$ and an MC simulation of the thermalization of electrons through the ``10-100 eV gap''. We find that 100-eV electrons thermalize in $\sim0.4$ and $\sim1.0$~ps for C and $\beta$-Ga$_2$O$_3$, respectively, while generated holes thermalize in $\sim0.5$~ps for both.  We also find that electrons travel a distance of order 100~nm during the first picosecond and that their spread is approximately isotropic in C and strongly anisotropic in $\beta$-Ga$_2$O$_3$. Lastly, we calculate EHP creation energies of 12.87 and 11.24~eV for C and $\beta$-Ga$_2$O$_3$, respectively.

We would like to acknowledge C. G. Van de Walle, S. T. Pantelides, R. D. Schrimpf, and D. M. Fleetwood for their support and especially for their contributions to our work on GaN. We acknowledge also the computational support of the Texas Advanced Computing Center (TACC) at The University of Texas at Austin for providing high-performance computing. This work has been supported by the Air Force Office of Scientific Research: Award FA9550-23-1-0549, and through the Center of Excellence in Radiation Effects: Award FA9550-22-1-0012.

The authors have no conflicts to disclose. The data that support the findings of this study are available from the corresponding author upon reasonable request.

{\bf Dallin Nielsen:} Conceptualization (supporting); formal analysis (lead); methodology (supporting); software (lead); writing - original draft (lead); writing - review and editing (equal); investigation (lead). {\bf Massimo Fischetti:} Conceptualization (lead); supervision (lead); methodology (lead); formal analysis (supporting); writing - review and editing (equal); resources (lead). 

% Create the reference section using BibTeX:
\bibliography{APL_don}

\end{document}